\documentclass[final]{svjour3}
\usepackage{graphicx}
\usepackage{rotating}
\usepackage{amsmath,mathtools}
\usepackage{amssymb}
\usepackage{mathptmx}
\makeatletter
\def\cl@chapter{\@elt {theorem}}
\makeatother
\usepackage{cleveref}
\usepackage[numbers]{natbib}
\makeatletter
\journalname{Journal of Low Temperature Physics}
\bibpunct{}{}{,}{s}{}{,}

\graphicspath{{fig/}}
\begin{document}

\title{A Model on Heat Signal of Crystal Detector at Low Temperature}

\author{Jin Li \and Inwook Kim}
\institute{ Center for Underground Physics, 
 Institute for Basic Science (IBS),
 Daejeon, 34047, Korea\\
  Tel.:+82-2-878-8524 \\
\email{jinlee@ibs.re.kr} %
}

\maketitle

\begin{abstract}
  We present a model to calculate heat signal shapes from low
  temperature bolometer attached to a crystal. 
  This model is based on the elementary
  acoustic wave theory at low temperature, and has been developed
  using modern Monte Carlo techniques. Physical processes in phonon
  propagation, such as transmission, scattering and reflection are considered.
  Using our model, the calculated time dependence of signal agrees with
  real experimental data.  This model has applications
  in low temperature rare event particle detectors for dark matter
  and neutrinos.


\end{abstract}

\section{Motivation}
With the development of low temperature detectors for rare event
searches, there is a growing
interest in understanding observed signals from basic physical
principles~\cite{HJMaris93,Brandt:2012zzb}.
Previous attempts are either limited in time scale or in type
of physics processes considered. Thus a new way to quantitatively
describe the observables is required.

Here we study signal on solid state crystals, which are widely used
to detect energy deposit to it by elementary particles.
Phonons are created by initial energy deposit, and then transported
to the sensor for detection.  The physical principles governing
the whole process are known to a large extent~\cite{SWLeman12}.
However, it remains to be accomplished to explain and predict
the observed signals using those principles.

\section{Physical Processes for Phonon}
\begin{figure}
\includegraphics[width=0.5\textwidth]{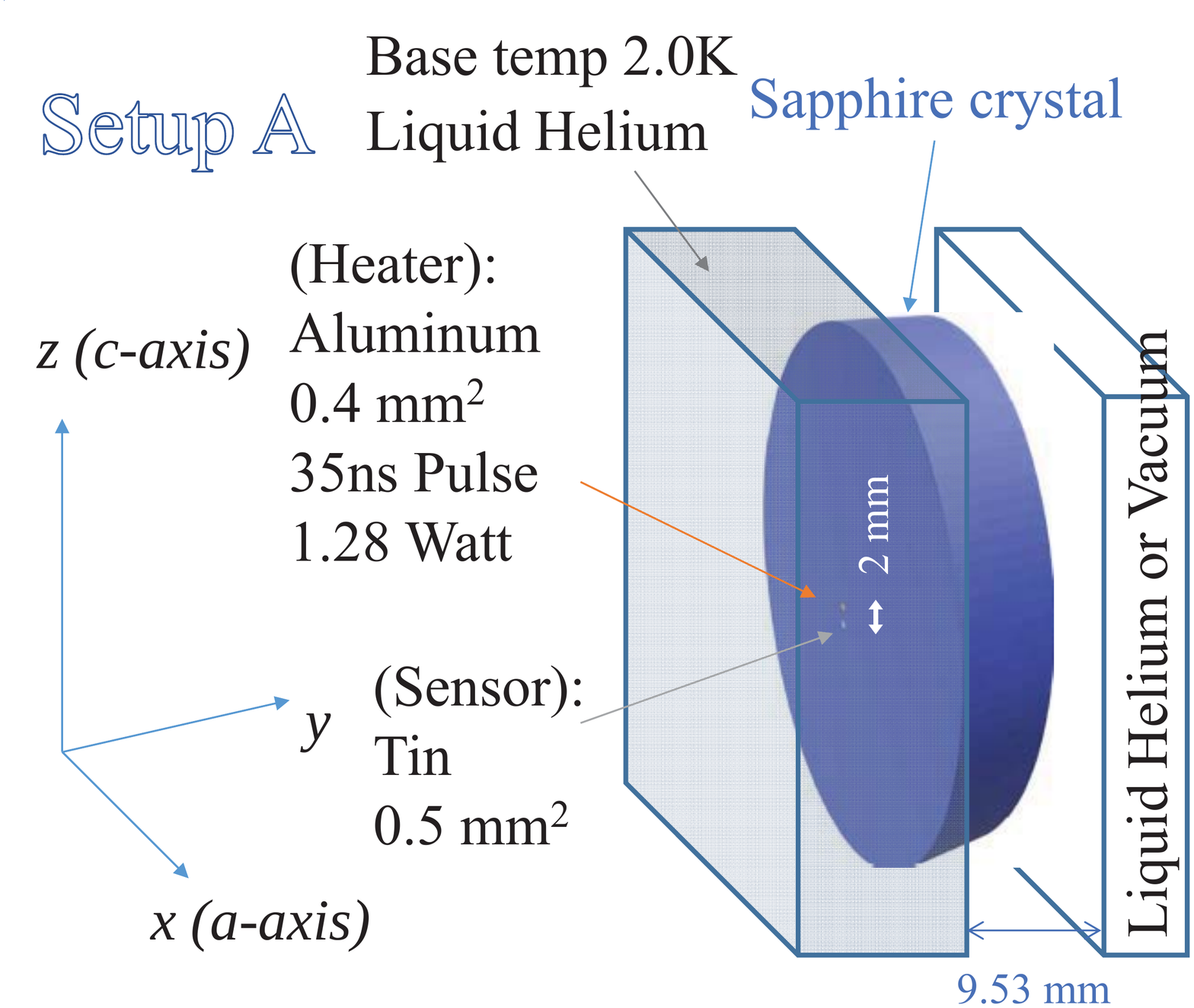}%
\includegraphics[width=0.5\textwidth]{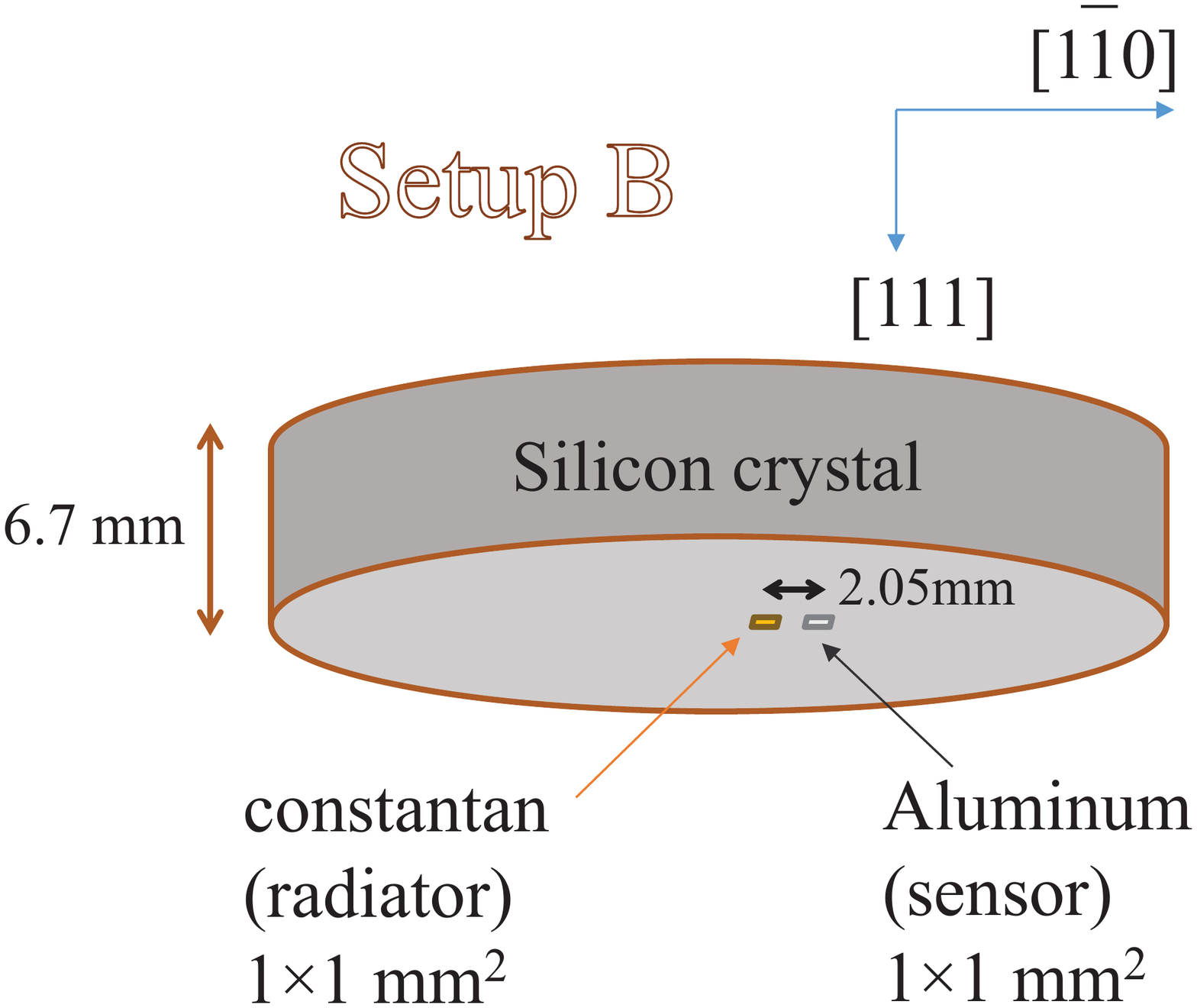}%
\caption{Experimental setup A (left) and setup B (right). The dimension
of the radiator, sensor and crystal substrate are shown.}
\label{fig:geom}
\end{figure}

In this study, we shall apply our model to experiments where
where transmission, scattering and reflection processes all play
important roles in the obtained data.
We choose detector geometry where
the phonons are emitted and detected from two small metal films
evaporated to the same cut surface of a crystal substrate,
such as two setups shown in Fig.~\ref{fig:geom}.
Here, in addition to free propagation,
the detected phonons must undergo either scattering in the bulk
or reflection from the surface of the other side, because
the solid angle of a straight line from the radiator source
to the sensor is zero.
This choice of detector geometry
provides a rigorous test to all physical processes considered
in our model.  

For the detector setups we study in this paper,
the phonons are initially created
in the radiator which is an ohmic metal heater, then transmit across
the interface to the crystal substrate, travel inside it, and
are finally absorbed by hitting
a superconducting bolometer with metal sensor.

In order to obtain the energy deposit spectrum,
we generate individual phonons, and simulate their transportation
from initial creation to the final absorption by applying
basic physical principles.
The physical processes used are discussed below.

\subsection{Phonon emission from the radiator to the crystal substrate} 
%\begin{itemize} %[leftmargin=*]
Initially, phonons are created in the radiator by ohmic heating, and then
transmit in to the crystal through the boundary.
The metal radiator is normally of polycrystalline nature, which can be
described as an isotropic material for acoustic properties.
The phonons in the radiator follow
the Planck distribution of energy levels.  The area of the radiator
$A^{(1)}$, the time-dependent temperature $T^{(1)}(t)$ of the radiator (1)
and base temperature $T^{(0)}$ of crystal (0), are firstly denoted.
Then the spectral power of the heat transmission to the crystal
$P^{(10)}(\omega,t)$ is~\cite{weis69}:
\begin{equation}
P^{(10)}(\omega,t) = 
\begin{dcases} A^{(1)}\frac{\hbar\omega^3}{8\pi^2}
\sum_{\sigma'=1}^{3} \frac{e_{\sigma'}^{(10)}}{{\{c_{\sigma'}^{(1)}\}}^2}
\cdot \left\{ \frac{1}{e^{\hbar\omega/k_BT^{(1)}(t)}-1} - 
  \frac{1}{e^{\hbar\omega/k_BT^{(0)}}-1} \right\} & \omega<\omega_{\max,\sigma'}\\
0 & \omega>\omega_{\max,\sigma'}
\end{dcases} \label{eqn:P10t}
\end{equation}
The halfspace emissivities $e_{\sigma'}^{(10)}$ from radiator (1) to
crystal (0)
can be calculated using the acoustic mismatch model as in Ref.~\cite{Rosch77}.
$c_{\sigma'}^{(1)}$ are the isotropic sound velocities for mode $\sigma'$
in the radiator.

In cases discussed in this study, cutoff frequency $\omega_{\max}$
can be considered to be big. 
By integrating Eq.~\ref{eqn:P10t} over frequency $\omega$,
the total transmitted power is related to $T^{(1)}(t)$:
\begin{equation}
P^{(10)}(T^{1}(t),T^{(0)}) = A^{(1)}\cdot \frac{\pi^2 k_B^4}{120\hbar^3}
\sum_{\sigma'=1}^{3}\frac{e_{\sigma'}^{(10)}}{{\{c_{\sigma'}^{(1)}\}}^2}
\cdot \left[ \left\{T^{(1)}(t)\right\}^4 - \left\{T^{(0)}\right\}^4 \right].
\label{eqn:temp}
\end{equation}
As an ohmic heat pulse is applied to the radiator, it quickly comes to 
a steady-state in less than a few ns for a typical thickness
of 50 nm~\cite{bayrle89}.  For the heat pulse longer than tens of ns,
we can use a steady-state radiator temperature $T^{(1)}$,
calculated by equaling the $P^{(10)}(T^{(1)},T^{(0)})$ to the input
electric power.

The differential distribution of the transmitted phonons into the
crystal substrate originating from the radiator-crystal interface
as a function of the wave vector $\vec q$ needs to be known to 
generate phonon directions in half-space.
The $\vec{q}$-space source distribution is obtained by differential
emissivities from acoustic boundary condition.  Let
the normal of the substrate surface pointing inward be $\vec e_3$,
the group velocity of mode $\sigma$ in substrate be
$\vec w_\sigma^{(0)}(\vec q)$,
the phase velocity magnitude of mode $\sigma$ in substrate
be $c_\sigma^{(0)}(\vec q)$.
The differential intensity of the phonons transmitted into the substrate
as a function of $(\omega, t, \vec{q})$ and mode $\sigma$,
is then~\cite{Muller90}:
\begin{equation}
i_\sigma^{(0)}(\omega,t,\vec{q})
= A^{(1)}\frac{\hbar\omega^3}{8\pi^3}\cdot
\frac{\vec e_3\cdot\vec w_\sigma^{(0)}(\vec q)}{\{c_\sigma^{(0)}(\vec q)\}^3}
\cdot \left\{ \frac{1}{e^{\hbar\omega/k_BT^{(1)}(t)}-1} - 
  \frac{1}{e^{\hbar\omega/k_BT^{(0)}}-1} \right\}
\cdot \sum_{\sigma'=1}^{3}t_{\sigma\sigma'}^{(01)}(-\vec q).
\label{eqn:i0diff}
\end{equation}
Here the differential transmittivity~\cite{Weis79}
$t_{\sigma\sigma'}^{(01)}(-\vec q)$
is used, with the sign
of $\vec q$ inverted, because it should be calculated in the opposite direction
assuming the wave is coming from the crystal and transmitted  to the radiator.

\subsection{Phonon transportation in the substrate}

A phonon quantum with frequency $\omega$ travels in the direction of
its group velocity, carrying an energy of $\hbar\omega$.  According to
the polarization vector, phonons belong to one of longitudinal (L),
fast transverse (FT) and slow transverse (ST) modes~\cite{Wolfe98}.
The propagation law of acoustic waves are governed by the elastic
constants $C$ and the density $\rho$ of the crystal. 
Written in component form, the group velocity $\vec w$ are related with the
wave vector $\vec q$ as: $w_n = (1/\rho\omega)C_{ijln}e_iq_je_l$.  The
normalized polarization vector $\vec e$ is determined by the wave equation
$\rho\omega^2e_i = C_{ijlm}q_jq_me_l$.

Mass defects in the lattice lead to scattering of phonons.
For a certain type of atom impurity $a$ with mass difference $\Delta m_a$
compared to the average atomic mass, the molecule mass $m$,
the probability of scattering per unit time to a final mode $f$ 
within a solid angle range $\Delta\Omega_f$ 
in wave vector $\vec q$ space is~\cite{Weis95}:
\begin{equation}
\Gamma_a =\frac{1}{16\pi^2}\frac{m}{\rho}f_a\left(\frac{\Delta m_a}{m}\right)^2
\omega^4(\vec e_i\cdot\vec e_f)^2(\Delta\Omega_f/c_f^3)
\label{eqn:sca}
\end{equation}
Here $f_a$ is number density concentration, calculated as
the numbers of impurity atoms $a$ divided by the total number of molecules.
$\vec e_i$ and $\vec e_f$ are the polarization vectors of the initial
and final states. 
$c_f$ is the phase velocity of the final state.
For several impurities, total scattering rate is obtained
by summing all impurity type $a$ and solid angles:
 $\Gamma = \sum_a\int d\Omega_f \Gamma_a $.  
Let $w$ be group velocity magnitude of the initial phonon,
then mean free path is $\lambda = w/\Gamma$.

On the substrate-vacuum boundary, one phonon quantum reflects with certain
probabilities to one of the three modes. 
For a boundary surface
with the normal at direction $\vec z$ pointing into the crystal medium
and an incident with wavevector $\vec q$,
the wavevector component parallel to the surface is $\vec q_\parallel = 
\vec q - (\vec q \cdot \vec z)$.  The Snell's law requires
$\vec q_\parallel$ to be equal for incident and reflected waves.
In general, there are three reflected wave modes for one incident wave.

The amplitudes for specular reflection waves are related to the incident waves
by requiring the vanishing of the stress vector across
the surface~\cite{Fedorov68}, because
it is a reflection at a stress-free boundary, with the other side being vacuum.
While the diffusive reflection is possible, it is not important in
experimental setups discussed in this paper
and is not considered in physics modeling.

\subsection{Phonon absorption in the sensor}
The incoming phonons which hit the sensor after transportation
are not in equilibrium, so each one needs to be treated individually.
The total energy deposit of a phonon to the sensor can be calculated
as a process of consecutive energy loss of phonon
interacting with electrons in a thick medium~\cite{Goetze76}.

The transmission probability $t^{(02)}_\sigma$ for normal incidence
and mode $\sigma$ 
depends on the mass densities $\rho^{(\ )}$ and phase velocities $c^{(\ )}$
for the substrate (0) and sensor (2), as:
\begin{equation}
t^{(02)}_\sigma = 1- \left[\frac{\rho^{(0)}c_\sigma^{(0)}-\rho^{(2)}c_\sigma^{(2)}}
               {\rho^{(0)}c_\sigma^{(0)}+\rho^{(2)}c_\sigma^{(2)}}\right]^2.
\end{equation}
In addition to the initial transmission, 
a phonon entering the sensor undergoes attenuation
in the metal and multiple reflections along the interfaces.
As a result, the total energy deposit
is the incident phonon energy scaled by an absorption
coefficient for sensor material~\cite{Mrzyglod94}:
%rev2
\begin{equation}
 a_\sigma^{(02)}(\omega) = t^{(02)}_\sigma\cdot
 \frac{1-e^{-4\alpha_\sigma^{(2)}(\omega)l^{(2)}}}
 {1-(1-t_\sigma^{(02)})e^{-4\alpha_\sigma^{(2)}(\omega)l^{(2)}}}
\label{eqn:acoeff}
\end{equation}
Here the absorption constant $\alpha_\sigma^{(2)}(\omega)=
\alpha_\sigma^{(2)}(\omega_0)\cdot \omega/\omega_0$ for
mode $\sigma$ is proportional to the phonon frequency $\omega$.
$l^{(2)}$ is the sensor film thickness.

By accumulating large amount phonons hitting at sensor at different times,
the input power as a function of time is obtained.  Using Eq.~\ref{eqn:temp}
by replacing the material superscript $(1)$ to $(2)$,
the temperature rise as a function of time can be
subsequently calculated.

\section{Simulation Procedure}
The actual simulation follows the physics model presented at the previous sections.  Phonon tracks are monitored from the creation until its absorption in
the sensor using modern Monte Carlo techniques. 
The detailed procedures for one phonon track are:
\begin{itemize}
\item Generate one phonon track traveling into the halfspace of crystal.
The initial vertex is evenly distributed on
the radiator-crystal interface area.
 The mode, direction and energy 
follow the distribution in Eq.~\ref{eqn:i0diff}.
Here the base temperature $T^{(0)}$ for substrate 
follows experimental condition, and the temperature for the radiator
$T^{(1)}$ is obtained from Eq.~\ref{eqn:temp},
assuming steady condition with a constant input power $P^{(10)}$.

\item Propagate the phonon track inside the crystal substrate.  The propagation is simulated with many connected tracks, with the direction following its group velocity.  The length of one individual track is determined by sampling the exponential distribution with the mean free path as its mean value.  The scattering process is implemented by generating a new track,
with the direction in $\vec q$ space and mode $f$ distributed 
in Eq.~\ref{eqn:sca}.
If the length of a track runs beyond the boundary of the substrate,
then it is reflected.  The mode of reflection is generated 
with a probability proportional to the power transmission across the interface.
If the track hits the area of the sensor, 
it is removed from the simulation and the energy deposit is 
calculated as in the absorption process of the next step.
\item Phonon absorption.  When a phonon track hits the boundary between
the sensor and the crystal,
the energy deposit to the sensor is calculated as the phonon energy
multiplied by the absorption coefficient $a_\sigma^{(02)}(\omega)$ 
in Eq.~\ref{eqn:acoeff}.
\end{itemize}

The starting time of a phonon track from the radiator is uniformly
sampled in the window of the heater pulse width.
To proceed a full simulation, huge number of phonon tracks
are generated and propagated, such that we
can obtain a smooth distribution of energy deposit versus time for the sensor.
If absolute energy deposit or temperature rise in the sensor
is needed, the absolute y-axis value is scaled according to
the ratio of total generated phonon energy over total input energy during
the input pulse.

\section{Experiments}
A simulation can be claimed useful only if it can explain and predict the experimental data.  We use two experimental setups: A as in Ref.~\cite{Taborek80}
and B as in Ref.~\cite{Mrzyglod95} to perform the simulation.
The input conditions are exactly the same as in the referenced papers.
For setup A, the material for radiator, substrate, and
sensor are aluminum, $m$-cut sapphire crystal and tin. 
For setup B, the corresponding materials are constantan, $[111]$-cut
silicon, and aluminum.
The geometries for two setups are illustrated in Fig.~\ref{fig:geom}.
The thickness of sensor are 200 (50) nm  for setup A (B).
The base temperatures are 2.0 K for setup A, and 1.369 K or 1.36 K for two
experiments in setup B.

\begin{figure}\centering
\hspace{-1.5pt}%
\includegraphics[width=0.56\textwidth]{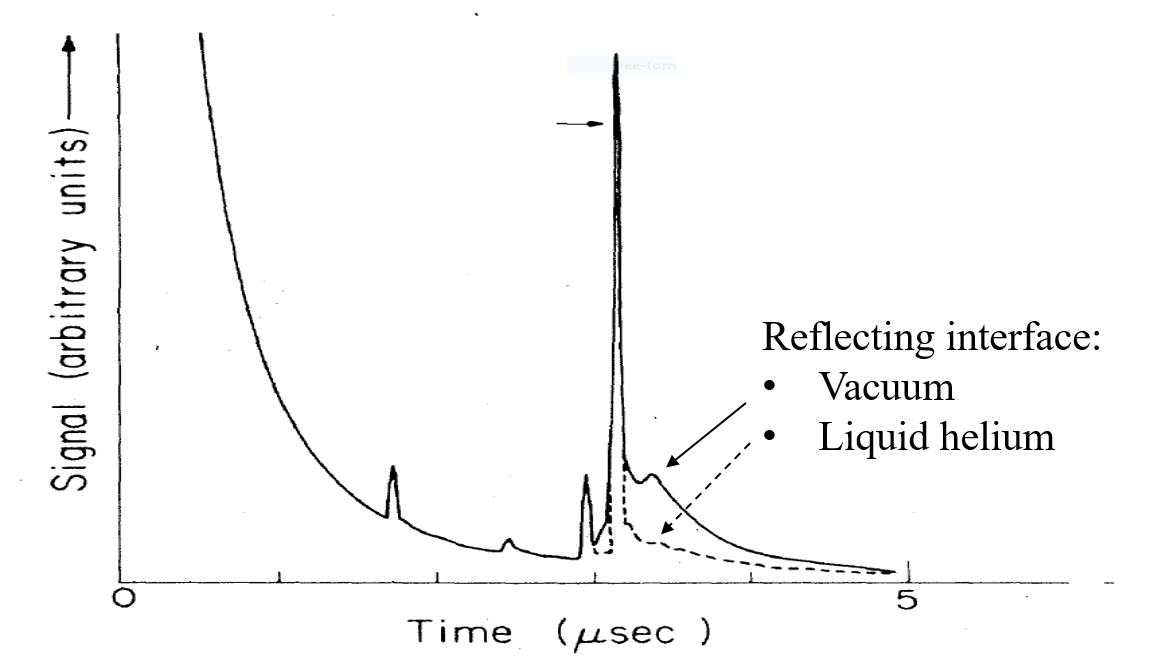}\\
\includegraphics[width=0.55\textwidth]{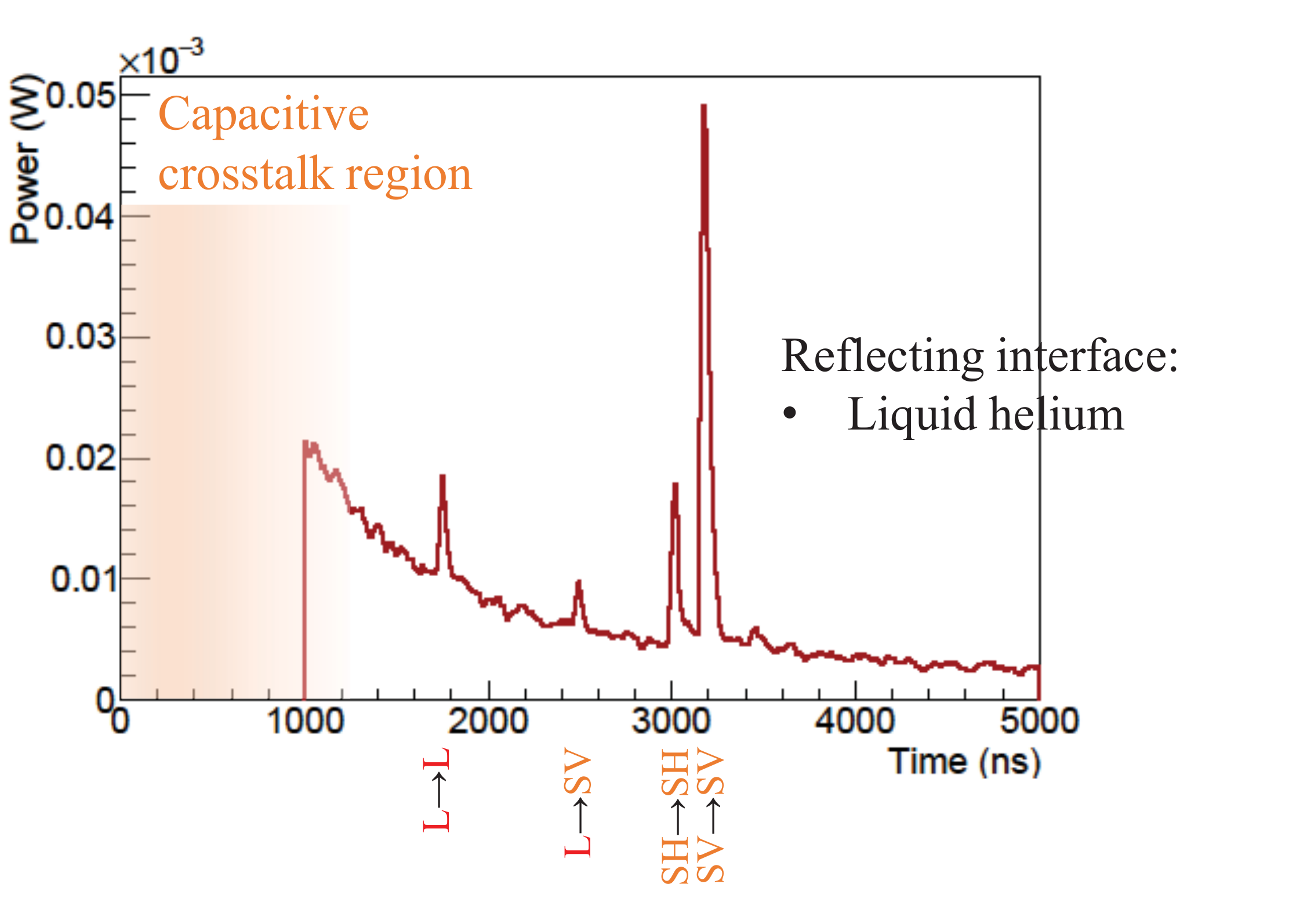}%
\caption{ Spectra of energy deposit for Experimental setup A
with a heater pulse width of 35 ns.
The top plot is for real experimental data from Taborek
and Goodstein~\cite{Taborek80}
and the bottom plot is for the simulation result.
The simulation is under a condition with the reflection surface in contact
with liquid helium, and should be compared with dashed line in the top plot.}
\label{fig:eSimExpA}
\end{figure}

The surface with radiator and sensor film in setup A
is immersed in liquid helium instead of vacuum.
The effect of helium contact to radiator and sensor is considered
by assuming that 50\% of phonons~\cite{Timothy11}
travel from the metal to the liquid helium at the interface,
which largely gives an overall scale change to the total emission
and absorption powers.
Because the experimental result of Setup A is a plot with
arbitrary unit in signal height, this effect is not relevant.
The reflection surface in setup A is kept in vacuum or liquid helium.
From the study in Ref.~\cite{Taborek80},
if the reflecting surface is in contact with liquid helium, 
no energy leak will happen for the specular reflection,
while the diffusely scattered phonons are lost due to the 
anomalous Kapitza effect.
Since the diffusive scattering process are not considered
in our simulation, we shall use the
experimental data with reflection surface in liquid helium
in comparison
(the dashed line in the top plot of Fig.~\ref{fig:eSimExpA}).

\begin{figure}
\includegraphics[width=0.5\textwidth]{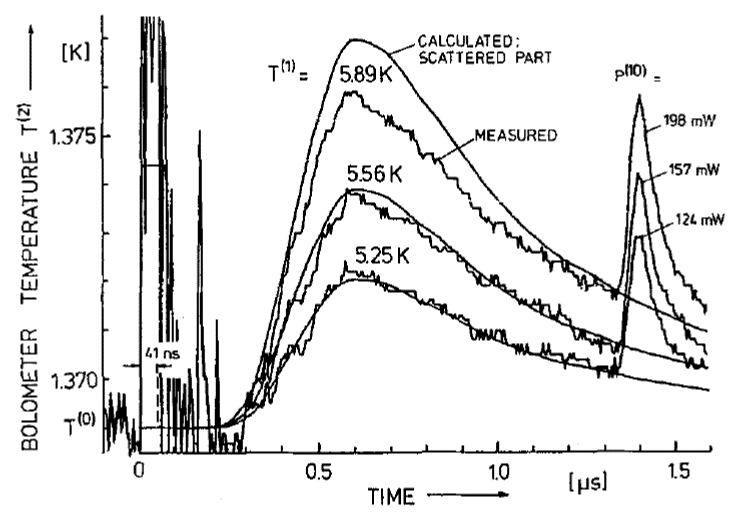}%
\includegraphics[width=0.5\textwidth]{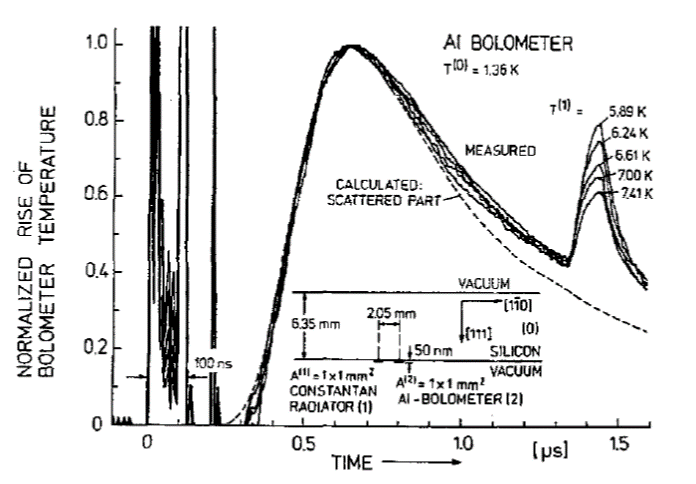}\\
\includegraphics[width=0.5\textwidth]{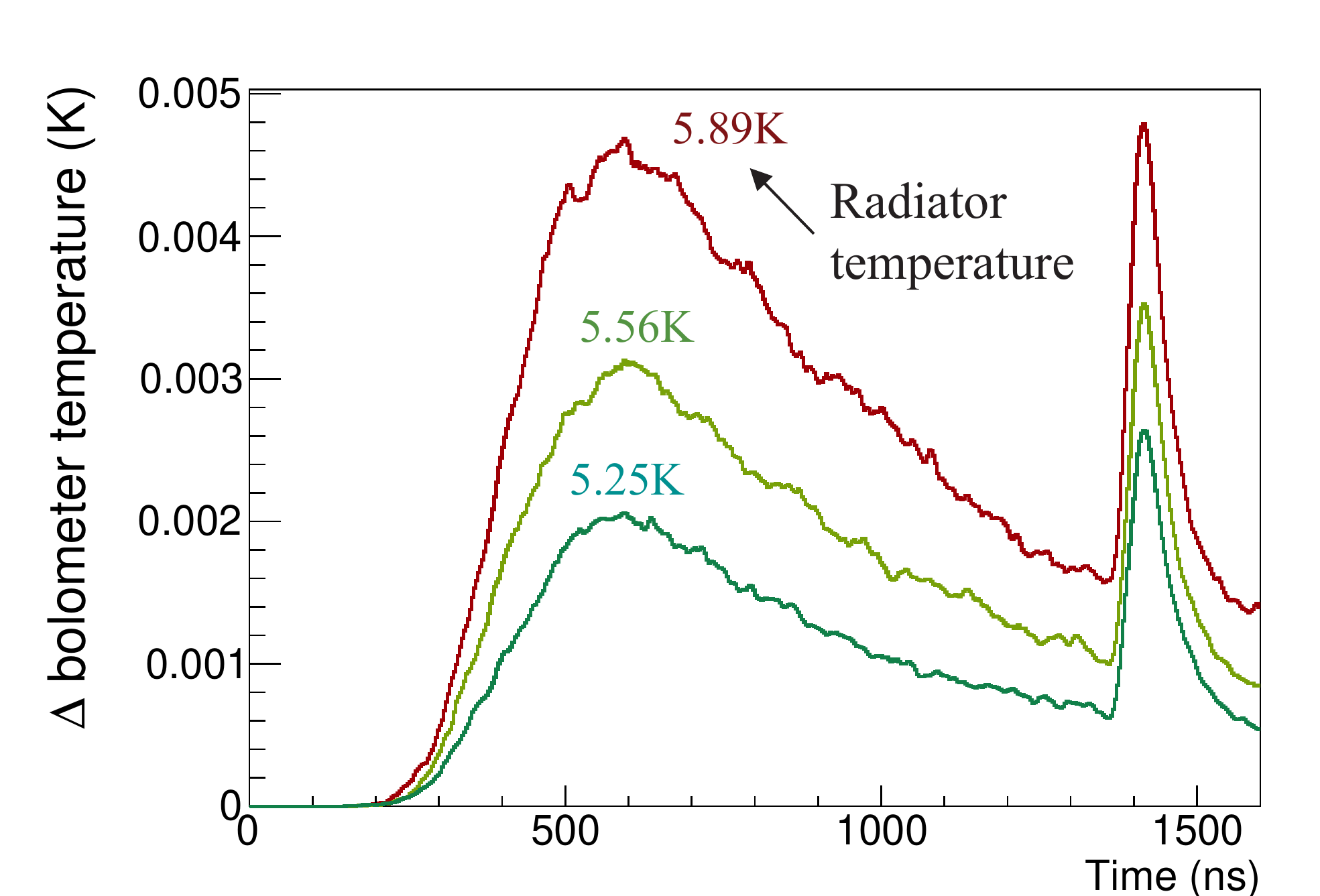}%
\includegraphics[width=0.5\textwidth]{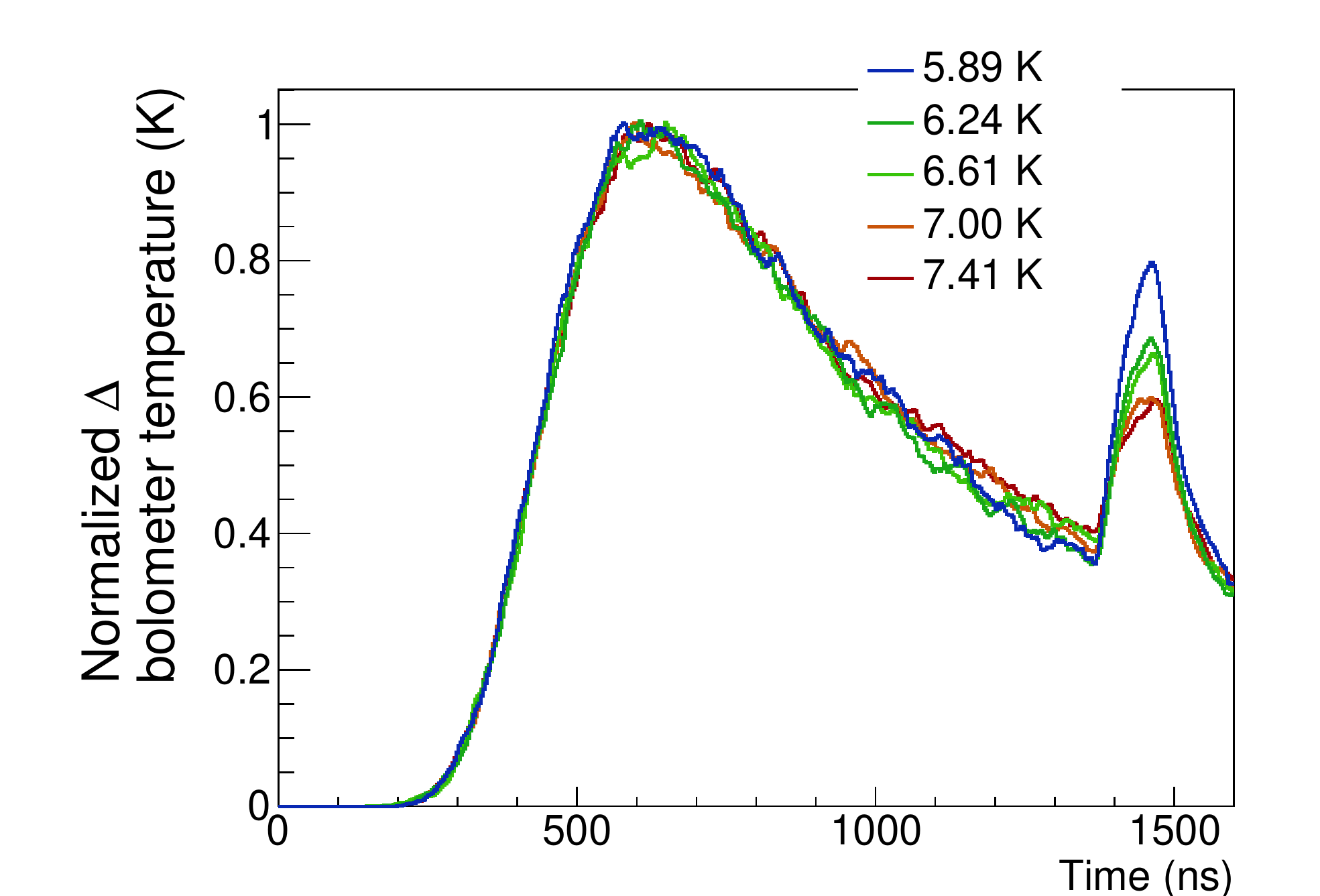}%
\caption{ Spectra of temperature rise for setup B.
The heater pulse width is 41 ns (left) and 100 ns (right).
The spectra labeled as ``MEASURED'' in the
top row is for real experimental data from 
Mrzyglod and Weis~\cite{Mrzyglod95}
and the bottom row is for the simulation result. Results with 
different radiator temperatures are plotted together. 
The right column
shows temperature rise normalized to same maximum height.  }
\label{fig:eSimExpB}
\end{figure}

In the simulation,
the elastic constants of the crystals  are obtained from
Ref.~\cite{AGEvery84}  for sapphire and Ref.~\cite{Mrzyglod95} for silicon.
The radiator's elastic constants are calculated from the longitudinal
and transverse velocities in Ref.~\cite{Rosch77}.  
The absorption coefficient $\alpha_\sigma^{(2)}(100 \mathrm{GHz})$
used for the aluminum and tin metal sensor
are measured values, as in Table I of Ref.~\cite{Mrzyglod94}.

A total of 100 million phonons are generated for setup A.  And
for each experimental conditions in setup B, 40 million phonons are generated.
The simulated timing spectrum of the energy deposit or temperature rise
in the sensor are shown and compared to the experimental data in
Figs.~\ref{fig:eSimExpA} and \ref{fig:eSimExpB}.
The time windows are chosen according to the published plots
of real data.  All plots show one or several peaks due to reflection
lying on a broad shape due to scattering.

In Fig.~\ref{fig:eSimExpA} for setup A,
the time window is wide enough to see various peaks.
Since the radiator-sensor is in the $C-Y$ plane, the transverse modes
can also be labeled as shear horizontal (SH) and shear vertical (SV) too.
The four distinct peaks with various intensity
correspond to the reflections of
L$\to$L, L$\to$SV, SH$\to$SH and SV$\to$SV transitions.  Other
possible reflections are not seen due to limited wave vector phase space or 
low reflection probabilities.

For setup B,
in the left column of Fig.~\ref{fig:eSimExpB}
three experimental conditions are applied
with different heater pulse power,
resulting the temperature of the radiator being $5.25$,
$5.56$ and $5.89$ Kelvin.  The width of time window is set to be just beyond
the first clear L$\to$ L reflection peak at around 1450 ns.
Broad bumps before the reflection peaks are due to bulk scattering, which
have maxima at around 600 ns from geometry effect of scattering vertices.
In the right column of Fig.~\ref{fig:eSimExpB},
superimposed plots are normalized to same maximum
height, with five different experiments with various radiator temperatures.
Here the relative size of reflection peaks decreases as the
 radiator temperature rises, which is due to the scattering probability
being proportional to the fourth power of the energy of individual phonons.
As phonon energy increases, it will encounter more scattering
against ballistic reflection.

We see an excellent agreement between the simulation
and real data in Fig.~\ref{fig:eSimExpA} for setup A,
and in left and right columns of Fig.~\ref{fig:eSimExpB}
for three and five different setup B conditions respectively.
This validates our model, at least for the metal heater
in a substrate geometry, where the initial heat pulse arrivals to
the sensor with reflection.

Previous attempts in Refs.~\cite{Taborek79,Muller90,Muller90_1}
have tried to explain the observed signal shape and magnitude.
Those methods used numerical iteration to determine all possible
phonon reflection paths, and numerical or analytical integration over
all the scattered phonon travel paths, by
treating the radiator and sensor as single points.
They are limited to time range of the first reflection,
single scattering in weak scattering limit, small size radiator and sensor,
thin geometry and can not extend to other applications straightforwardly.
Moreover, in Ref.~\cite{Taborek79}, a quantitative explanation
of the size of each peak cannot be obtained
for the multiple-peak spectrum in Setup A.
In a simulation reported on Ref.~\cite{Brandt:2012zzb},
no reflection process is considered,
no full anisotropic treatment of scattering is done,
and no rigorous check has been made for different experimental conditions.
On the contrary, with the full simulation method presented in this work,
none of the above limits exist.
In addition, we are free to extend with other physical processes
which may be important in different experimental conditions.
It is clear that simulation methods similar to our model
must be done to explain the experimental data quantitatively
 in real applications.

The next step is to apply our model to larger detectors
and longer collection time.  Examples are
cryogenic crystal detectors for rare event search experiments,
such as CaMoO$_4$ based neutrinoless double beta decay search
AMoRE~\cite{GBKim15}
and CaWO$_4$ based dark matter search CRESST~\cite{Angloher:2015ewa}.
It is of importance to understand
phonon creation process by energy deposit from elementary particles.
Also, additional phonon propagation process may need to be considered,
for instance, the diffusive scattering at the boundary, which can affect
the energy transportation at longer time scale.

\section{Conclusion and Discussion}
A new model has connected the real signals of low temperature crystal
and the underlying first principle physical processes. It is a result of
combining the evolving computing technology and the basic theory of
acoustic waves. Thus, from now, our community has a new tool to
explain and predict full series of data spectrum quantitatively.
This tool is expected to be useful when applying to larger and more
complex experimental setups. Here, new questions in phonon physics
may appear in the development of phonon-based detectors.

\begin{acknowledgements}
The author would like to thank Yonghamb Kim, Juhee Lee, Chang Lee
and Seungyoon Oh for useful communications.  This work was supported by
project code IBS-R016-D1.
\end{acknowledgements}
%%\pagebreak


\begin{thebibliography}{99}

\bibitem{HJMaris93}
H.~J.~Maris, {\it J. Low Temp. Phys.} \textbf{93}, 355, (1993).

\bibitem{Brandt:2012zzb}  %CDMS-Ge
  D.~Brandt, M.~Asai, P.L.~Brink et al.,
  %``Monte Carlo Simulation of Massive Absorbers for Cryogenic Calorimeters,''
  {\it J.\ Low.\ Temp.\ Phys.}  \textbf{167}, 485, (2012).
%doi:10.1007/s10909-012-0480-3

\bibitem{SWLeman12}
Steven W. Leman, {\it Review of Scientific Instruments} \textbf{83},
091101, (2012).

\bibitem{weis69}
O.~Weis, {\it Z. Angew. Phys.} \textbf{26}, 325, (1969).

\bibitem{Rosch77}
F.~R{\"o}sch and O.~Weis, {\it Z. Phys. B} \textbf{27}, 33, (1977).

\bibitem{bayrle89}
R.~Bayrle and O.~Weis, {\it J. Low Temp. Phys.} \textbf{76}, 129, (1989).

\bibitem{Muller90}
G.~M{\"u}ller and O.~Weis, {\it Z. Phys. B} \textbf{80}, 25, (1990).

\bibitem{Weis79}
O.~Weis, {\it Z. Phys. B} \textbf{34}, 55, (1979).

\bibitem{Wolfe98}
J.P.~Wolfe, Imaging Phonons, Cambridge Univ. Press, Cambridge, 1998.

\bibitem{Weis95}
O.~Weis, {\it Z. Phys. B} \textbf{96}, 525, (1995).

\bibitem{Fedorov68}
F.~I.~Fedorov, {\it Theory of Elastic Waves in Crystals}, Ch.~8, Plenum,
New York, 1968.

\bibitem{Goetze76}
M.~Goetze, M.~Nover and O.~Weis, {\it Z. Phys. B} \textbf{25}, 1, (1976).

\bibitem{Mrzyglod94}
A.~Mrzyglod and O.~Weis, {\it J. Low Temp. Phys.} \textbf{97}, 275, (1994).

%experiment:
\bibitem{Taborek80}%setup A
P.~Taborek and D.~Goodstein, {\it Phys. Rev. B} \textbf{22}, 1550, (1980).

\bibitem{Mrzyglod95}%setup B
A.~Mrzyglod and O.~Weis, {\it Z. Phys. B} \textbf{97}, 103--112, (1995).

\bibitem{Timothy11} %LHe halfspace emission
T.~L.~Head and M.~E.~Msall, {\it Chinese Journal of Physics} \textbf{49}, 278, (2011).

\bibitem{AGEvery84} %sapphire elastic constants
A.~G.~Every and G.~L.~Koos and J.~P.~Wolfe, {\it Phys. Rev. B} \textbf{29}, 2190, (1984).

\bibitem{Taborek79}%setup A
P.~Taborek and D.~Goodstein, {\it J. Phys. C} \textbf{12}, 4737, (1979).

\bibitem{Muller90_1}
G.~M{\"u}ller and O.~Weis, {\it Z. Phys. B} \textbf{80}, 15, (1990).


\bibitem{GBKim15} %AMoRE experiment
G. B. Kim, S. Choi, F. A. Danevich, et al., {\it Advances in High Energy Physics}, vol. 2015, Article ID 817530, 7 pages, 2015.

\bibitem{Angloher:2015ewa} 
  G.~Angloher {\it et al.} [CRESST Collaboration],
  %``Results on light dark matter particles with a low-threshold CRESST-II detector,''
  Eur.\ Phys.\ J.\ C {\bf 76}, no. 1, 25 (2016).
%  doi:10.1140/epjc/s10052-016-3877-3
%  [arXiv:1509.01515 [astro-ph.CO]].

\end{thebibliography}
\end{document}